\documentclass[twocolumn]{elsart3p}
\usepackage[dvips]{graphicx}
\usepackage{amsmath}
\usepackage{amssymb}
\usepackage{times}
\newcommand{\bs}{\begin{split}}
\newcommand{\es}{\end{split}}
\newcommand{\bea}{\begin{eqnarray}}
\newcommand{\eea}{\end{eqnarray}}
\newcommand{\be}{\begin{equation}}
\newcommand{\ee}{\end{equation}}
\newcommand{\ba}{\begin{eqnarray}}
\newcommand{\ea}{\end{eqnarray}}

\newcommand{\uk}{u_{\mathbf{k}}}
\newcommand{\vk}{v_{\mathbf{k}}}

\begin{document}

\begin{frontmatter}

\title{What can ultracold Fermi gases teach us about high $T_c$
    superconductors and vice versa?}

\author{K. Levin and Qijin Chen}

\address{James Franck Institute and Department of Physics,
 University of Chicago, Chicago, Illinois 60637}

%\date{\today}

\begin{abstract}
  We review recent developments in the field of ultracold atomic Fermi
  gases. As the cold atom system evolves from BCS to Bose-Einstein
  condensation (BEC), the behavior of the thermodynamics, and the
  particle density profiles evolves smoothly in a way which can be well
  understood theoretically.  In the interesting ``unitary" regime, we
  show that these and other data necessarily requires the introduction
  of a pseudogap in the fermionic spectrum which exhibits many striking
  similarities to its counterpart in underdoped high $T_c$
  superconductors. We emphasize these similarities, giving an overview
  of the experimental tools and key issues of common interest in both
  systems.
\end{abstract}

\begin{keyword}
Ultracold Fermi gas \sep High $T_c$ superconductors \sep Fermionic
superfluidity \sep Feshbach resonance \hfill 
\textbf{\textsf{cond-mat/0611104}}

%\PACS 03.75.Hh \sep 03.75.Ss \sep 74.20.-z
\end{keyword}
\end{frontmatter}

The study of ultracold trapped fermionic gases is a rapidly exploding
subject \cite{JinGrimmKetterle,Thomas} which is defining new
directions in condensed matter and atomic physics.  It has also captured
the attention of physicists who study color superconducting quark matter
as well as nuclear matter.  Indeed, it is hard, in recent times, to find
a subfield of physics which appeals this broadly to the research
community.  What makes these gases (and lattices) so important is their
remarkable tunability and controllability. Using a Feshbach resonance,
one can tune the attractive two-body interaction from weak to strong,
and thereby make a smooth crossover from a BCS superfluid to a
Bose-Einstein condensation (BEC) \cite{Leggett}.  This allows high
transition temperatures $T_c$ (relative to the Fermi energy $E_F$) which
are interesting in their own right.  Importantly, they may also provide
insights into the high temperature cuprate superconductors
\cite{Chen2,ourreviews,LeggettNature}.

This paper will concentrate on those issues relating to the common
features of high $T_c$ and cold atom systems with particular emphasis on
the important ``pseudogap" effects. We will study BCS-BEC crossover in
atomic Fermi gases, looking at a number of different experiments which
suggest the existence of noncondensed pairs below $T_c$ and, their
counterpart, pre-formed pairs, above $T_c$. The latter are associated
with the fermionic pseudogap.  While there is much controversy in the
field of high $T_c$ superconductivity, there is a school of thought
\cite{ourreviews,LeggettNature,Uemura} which argues that these systems
are somewhere intermediate between BCS and BEC. As stated by A.~J.
Leggett: \textit{ ``The size of the [cuprate] pairs is somewhere in the
  range 10-30\,\AA\ -- from measurements of the upper critical field,
  Fermi velocity and $T_c$. This means that the pair size is only
  moderately greater than the inter-conduction electron in-plane
  spacing, putting us in the intermediate regime of the so-called
  Bose-Einstein condensate to BCS superconductor (BEC-BCS) crossover,
  and leading us to expect very large effects of fluctuations (they are
  indeed found)."}

Studies of BCS-BEC crossover are built around early observations by
Leggett \cite{Leggett} that the BCS ground
state, proposed by Bardeen, Cooper, and Schrieffer in 1957
\begin{equation}
\Psi_0=\Pi_{\bf k}(\uk+\vk c_{\bf k,\uparrow}^{\dagger}
c_{\bf -k,\downarrow}^{\dagger})|0\rangle \,,
\label{eq:1a}
\end{equation}
is much more general than was originally thought.  If one increases the
strength of the attraction and self-consistently solves for the
fermionic chemical potential $\mu$, this wave function will correspond
to a more BEC-like form of superfluidity.
Knowing the ground state what is the nature of the superfluidity at
finite $T$? That is the central question we will address in this article.

Even without a detailed theoretical framework we can make three
important observations.  (i) The fundamental statistical entities in
these superfluids are fermions. We measure the ``bosonic" or
pair-degrees of freedom indirectly via the fermionic gap parameter
$\Delta(T)$. This tells us about bosons indirectly through the binding
together of two fermions.
(ii) As we go from BCS to BEC, pairs will begin to form at a temperature
$T^*$ above the transition temperature $T_c$ at which they condense.
This pair formation is associated with a normal state pseudogap.
(iii) In general there will be two types of excitations in these BCS-BEC
crossover systems. Importantly in the intermediate case (often called
the ``unitary" regime) the excitations consist of a mix of both fermions
and bosons.  They are not independent since the gap in the fermionic
spectrum is related to the density of bosons.

\begin{figure}
%\centerline{\includegraphics[width=3.1in,clip]{ProfileDecomp.eps}}
\centerline{\includegraphics[width=3.1in,clip]{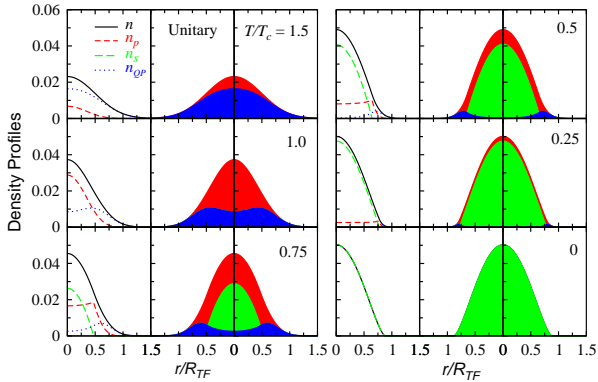}}
\caption{(Color online) Decomposition of density profiles at various $T$
  at unitarity. Here green (light gray) refers to the condensate, red
  (dark gray) to the noncondensed pairs and blue (black) to the excited
  fermionic states. $T_c = 0.27T_F$, and $R_{TF}$ is the Thomas-Fermi
  radius. 
%The presence of noncondensed pairs is essential \cite{JS5} for
%  explaining why there are no sharp features in these profiles,
%  associated with the interface of the normal and superfluid regions.
  Here $n_s$, $n_p$, and $n_{QP}$ denote density of superfluid,
  incoherent pairs, and fermionic quasiparticles, respectively.}
\label{fig:23new}
\end{figure}

To address $ T \neq 0$ we use a theoretical formalism based on a
$T$-matrix approach which is outlined in detail elsewhere
\cite{ourreviews}.  We turn now to experiments in cold atom systems
which help to establish the existence of noncondensed pairs, and the
related pseudogap.  In Fig. \ref{fig:23new} we plot a decomposition of
the particle density profiles \cite{JS5} in a trapped Fermi gas for
various temperatures above and below $T_c$.  The various color codes (or
gray scales) indicate the condensate along with the noncondensed pairs
and the fermions. This decomposition is based on the superfluid density
so that all atoms participate in the condensation at $T=0$.

\begin{figure}
%\centerline{\includegraphics[width=2.7in,clip]{AllT4.eps}}
\centerline{\includegraphics[width=2.7in,clip]{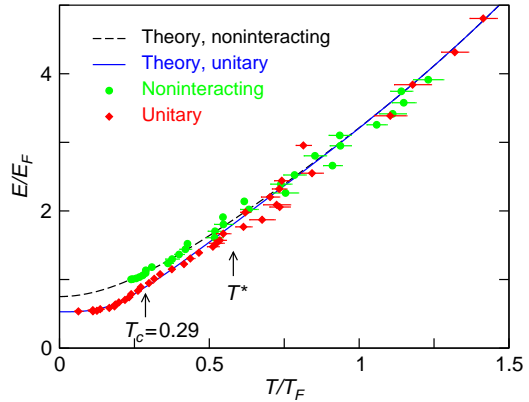}}
\caption{(Color online) Pseudogap effects as apparent from
  thermodynamics. Plotted is energy per particle $E/E_F$ versus $T/T_F$.
  From Ref.  \cite{ThermoScience2}. The fact that the experimental data
  (symbols) (and the two theoretical curves) for noninteracting and
  unitary Fermi gases do not merge until higher $T^* > T_c$ is
  consistent with the presence of a normal state pseudogap.}
\label{fig:232new}
\end{figure}

This figure contains a key point.  The noncondensed pairs are
responsible for smoothing out what otherwise would be a discontinuity
between the fermionic and condensate contributions.  This leads to a
featureless profile, in agreement with experiment \cite{Thomas}.
Indeed, these experimental observations originally presented a challenge
for previous theories which ignored noncondensed pair excitations, and
therefore predicted an effectively bimodal profile with a kink at the
edge of the superfluid core. One can see from the figure that even at
$T_c$, the system is different from a Fermi gas.  That is, noncondensed
pairs are present (in the central region of the trap) when the
condensate is gone.  Even at $T/T_c = 1.5$ there is a considerable
fraction of noncondensed pairs.
%the trap profile is not that of an ideal gas.
It is not until around $T^* \approx 2T_c$ for this unitary case, that
noncondensed pairs have finally disappeared.

\begin{figure}
%\centerline{\includegraphics[width=2.2in,clip]{Chin2.eps}}
\centerline{\includegraphics[width=2.2in,clip]{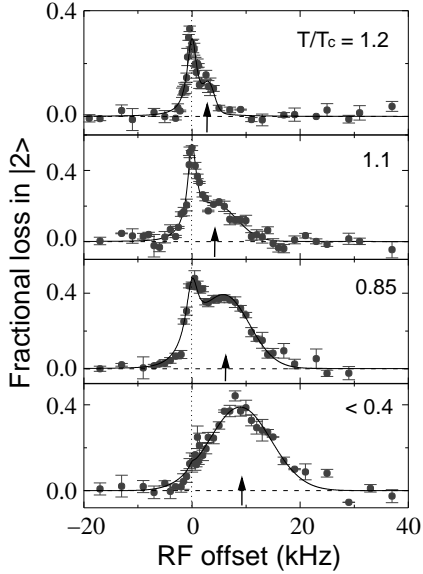}}
\caption{Experimental RF Spectra for $^6$Li at unitarity.
%  at 834\,G.  
  The temperatures labeled in the figure were computed theoretically at
  unitarity based on adiabatic sweeps from BEC.  The two top curves,
  thus, correspond to the normal phase, thereby, indicating pseudogap
  effects. Here $E_F = 2.5 \mu $K, or 52\,kHz.  From
  Ref.~\cite{Grimm4}.}
\label{fig:24Cheng}
\end{figure}

We next turn to a detailed comparison of theory and experiment for
thermodynamics.  Figure \ref{fig:232new} presents a plot of energy per
atom $E$ as a function of $T$ comparing the unitary and non-interacting
regimes.  The solid curves are theoretical while the data points are
measured in $^6$Li \cite{ThermoScience2}.  There has been a
recalibration of the experimental temperature scale
\cite{ThermoScience2} in order to plot theory and experiment in the same
figure.  The latter was determined via Thomas-Fermi fits to the density
profiles \cite{JS5}.  To arrive at the calibration, we applied the same
fits to the theoretically produced density profiles, examples of which
appear in Fig.~\ref{fig:23new}.

Good agreement between theory and experiment is apparent in
Fig.~\ref{fig:232new}.  In the figure, the temperature dependence of $E$
reflects primarily fermionic excitations at the edge of the trap
\cite{ChenThermo}, although there is a small bosonic contribution as
well.
Importantly one can see the effect of a pseudogap in the unitary case.
That is, the temperature $T^*$ is visible from the plots as that at
which the non-interacting and unitary curves merge.  This corresponds
roughly to $T^* \approx 2 T_c$.  In this way, this figure and the
proceeding figure are seen to be consistent.

Measurements \cite{Grimm4} of the excitation gap $\Delta$ can be made
more directly, and, in this way one can further probe the existence of a
pseudogap.  This pairing gap spectroscopy is based on using a third
atomic level, called $|3 \rangle$, which does not participate in the
superfluid pairing. Under application of radio frequency (RF) fields,
one component of the Cooper pairs, called $|2 \rangle$, is excited to
state $|3\rangle$.  If there is no gap $\Delta$ then the energy it takes
to excite $|2 \rangle$ to $|3 \rangle$ is the atomic level splitting
$\omega_{23}$. In the presence of pairing (either above or below $T_c$)
an extra energy associated with the gap $\Delta$ must be input to excite
the state $|2 \rangle$, as a result of the breaking of the pairs.
Figure \ref{fig:24Cheng} shows a plot of the spectra for $^6$Li near
unitarity for four different temperatures, which we discuss in more
detail below.  In general for this case, as well as for the BCS and BEC
limits, there are two peak structures which appear in the data and in
the theory \cite{Torma2,heyan}: the sharp peak at $\omega_{23} \equiv 0$
which is associated with ``free" fermions at the trap edge and the
broader peak which reflects the presence of paired atoms; more
precisely, this broad peak derives from the distribution of $\Delta$ in
the trap.  At high $T$ (compared to $\Delta$), only the sharp feature is
present, whereas at low $T$ only the broad feature remains.  The
sharpness of the free atom peak can be understood as coming from a large
phase space contribution associated with the $2 \rightarrow 3$
excitations \cite{heyan}.  These data alone do not directly indicate the
presence of superfluidity; rather they provide strong evidence for
pairing.

It is interesting to return to discuss the temperatures in the various
panels. What is measured experimentally are temperatures $T'$ which
correspond to the temperature at the start of an adiabatic sweep from
the BEC limit to unitarity. Here fits to the BEC-like profiles are used
to deduce $T'$ from the shape of the Gaussian tails in the trap. Based
on knowledge \cite{ChenThermo} about thermodynamics (energy $E$ in the
previous figure or, equivalently, entropy $S$), and given $T'$, one can
then compute the final temperature in the unitary regime, assuming $S$
is constant in a sweep.  We find that the four temperatures are as
indicated in the figures.  Importantly, one can conclude that the first
two cases correspond to a normal state, albeit not far above $T_c$.  In
this way, these figures suggest that a normal state pseudogap is present
as reflected by the broad shoulder above the narrow free atom peak.

\begin{figure}
%\centerline{\includegraphics[width=2.4in,clip]{phase.eps}}
\centerline{\includegraphics[width=2.4in,clip]{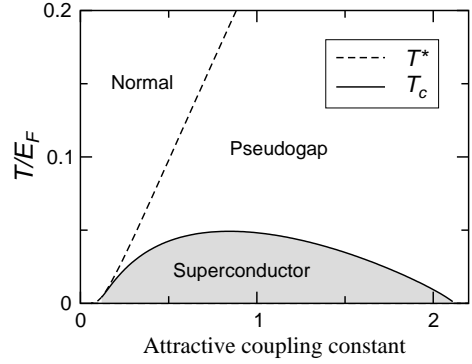}}
\caption{Phase Diagram for a quasi-two-dimensional $d$-wave
superconductor on a lattice. Here the horizontal axis corresponds to
the strength of attractive interaction
 $-U/4t$, where $t$ is the in-plane hopping matrix element.
}
\label{fig:23}
\end{figure}

\begin{figure}
%\centerline{\includegraphics[width=2.3in]{Broun_0509223v1_Fig2.eps}}
\centerline{\includegraphics[width=2.3in]{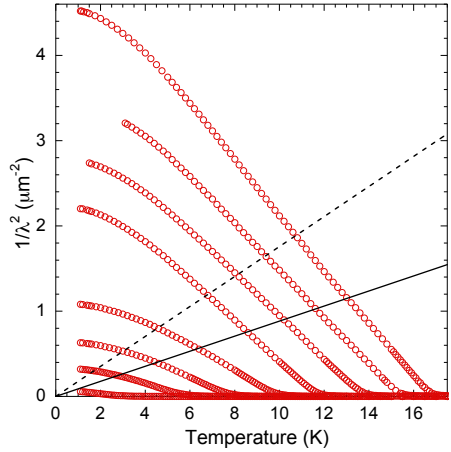}}
\caption{Experimental data from Ref.~\cite{Hardynew} on superfluid
  density in the extreme underdoped regime of the cuprate superconductors.}
\label{fig:5}
\end{figure}

We turn finally to studies of the BCS-BEC crossover scenario in the high
$T_c$ superconductors. The cuprates are different from the ultracold
fermionic superfluids in one key respect; they are $d$-wave
superconductors and their electronic dispersion is associated with a
quasi-two dimensional tight binding lattice. In many ways this is not a
profound difference from the perspective of BCS-BEC crossover.  Figure
\ref{fig:23} shows a plot of the two important temperatures $T_c$ and
$T^*$ as a function of increasing attractive coupling.  On the left is
BCS and the right is the intermediate or pseudogap (\textit{i.e.,} near
unitarity) regime. The BEC regime is not visible because $T_c$
disappears before it can be accessed.  This disappearance of $T_c$ is a
consequence of $d$-wave pairing which leads to pair localization effects
\cite{Chen1}.  If one chooses to fit the pairing temperature $T^*(x)$ to
experiment, this will relate the attractive coupling constant on the
horizontal axis directly to the hole concentration $x$.  The resulting
figure \cite{Chen1} is then very similar to its counterpart in the
cuprates.

\begin{figure}
%\centerline{\includegraphics[width=2.3in,clip]{Hc1_7.eps}}
\centerline{\includegraphics[width=2.3in,clip]{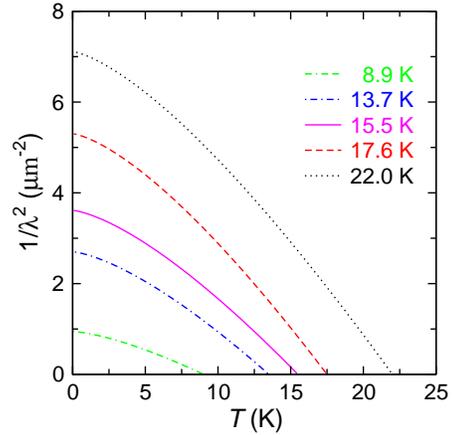}}
\caption{Theoretical calculation of superfluid density for underdoped
  cuprate superconductors, to be compared with Fig.~\ref{fig:5}.
}
\label{fig:6}
\end{figure}

To further probe the relevance of BCS-BEC crossover theory for the
cuprates we now address data below $T_c$ where noncondensed pairs are
predicted to be present in addition to the usual fermionic Bogoliubov
excitations.  Figure \ref{fig:5} presents experimental data
\cite{Hardynew} on the inverse square (in-plane) penetration depth
$1/\lambda^2(T)$ for a series of very underdoped cuprates with variable
hole stoichiometry or $x$.  What is striking is the fact that the curves
for each $x$ are all rather similar even though the fermionic gap (or
equivalently $T^*$) varies considerably for this range of $x$.  Thus,
one can conclude that the behavior of the superfluid density $n_s$
[$\propto 1/\lambda^2(T)$] does \textit{not} simply reflect the
fermionic gap energy scale.  This observation, which would be
paradoxical in strict BCS theory, is consistent with the picture that
fermionic degrees of freedom are not the only excitations of the
condensate to contribute to the superfluid density.  Bosonic excitations
are also present.  Figure \ref{fig:6} presents a plot of the theoretical
counterpart of $ 1/\lambda^2(T)$ within the crossover scenario. Here there
are bosonic as well as fermionic excitations of the condensate and the
calculations appear to be in good semi-quantitative agreement with the
data. The deviation from linearity at very low $T$ is a consequence of
incoherent pair excitations.

In conclusion, in this paper we have addressed commonalities of
ultracold trapped Fermi gases and high $T_c$ superconductors.  These
common features revolve around the scenario
\cite{LeggettNature,ourreviews} that the cuprates are somewhere
intermediate between BCS and BEC. Importantly, this scenario has been
directly realized in trapped Fermi gases. Here one sees considerable
evidence for pre-formed pairs and the related fermionic pseudogap which
appear reminiscent of their cuprate counterparts.

This work was supported by NSF PHY-0555325 and NSF-MRSEC Grant
No.~DMR-0213745.

\vspace*{-6mm}

%\bibliographystyle{prsty-varenna}

%\bibliography{Review2}

\begin{thebibliography}{10}
\setlength{\itemsep}{0.ex}

\bibitem{JinGrimmKetterle}
M. Greiner, C.~A. Regal,  D.~S. Jin, Nature (London) 426 (2003) 537; S.
  Jochim {\it et~al.}, Science 302 (2003) 2101; M.~W. Zwierlein {\it et~al.},
  Phys. Rev. Lett. 92 (2004) 120403.

\bibitem{Thomas}
K.~M. O'Hara {\it et~al.}, Science   298  (2002) 2179.

\bibitem{Leggett}
A.~J. Leggett,  in {\em Modern Trends in the Theory of Condensed Matter},
  edited by A. Pekalski and J. Przystawa (Springer-Verlag, Berlin, 1980), pp.\
  13--27.

\bibitem{Chen2}
Q.~J. Chen, I. Kosztin, B. Jank\'o,  K. Levin, Phys. Rev. Lett.  81
  (1998) 4708.

\bibitem{ourreviews}
Q.~J. Chen, J. Stajic, S.~N. Tan,  K. Levin, Phys. Rep. 412 (2005) 1; Q.~J.
  Chen, J. Stajic,  K. Levin, Low Temp. Phys. 32 (2006) 406.

\bibitem{LeggettNature}
A.~J. Leggett, Nature Physics 2  (2006) 134.

\bibitem{Uemura}
Y.~J. Uemura, Physica C 282-287  (1997) 194.

\bibitem{JS5}
J. Stajic, Q.~J. Chen,  K. Levin, Phys. Rev. Lett.  94  (2005)
  060401.

\bibitem{ThermoScience2}
J. Kinast, A. Turlapov, J.~E. Thomas, Q.~J. Chen, J. Stajic,  K. Levin,
  Science 307 (2005) 1296.

\bibitem{Grimm4}
C. Chin {\it et~al.}, Science 305  (2004) 1128.

\bibitem{ChenThermo}
Q.~J. Chen, J. Stajic,  K. Levin, Phys. Rev. Lett. 95  (2005)
  260405.

\bibitem{Torma2}
J. Kinnunen, M. Rodriguez,  P. T\"orm\"a, Science  305  (2004) 1131.

\bibitem{heyan}
Y. He, Q.~J. Chen,  K. Levin, Phys. Rev. A  72  (2005) 011602(R).

\bibitem{Hardynew}
D.~M. Broun {\it et~al.}, e-print cond-mat/0509223 (unpublished).

\bibitem{Chen1}
Q.~J. Chen, I. Kosztin, B. Jank\'o,  K. Levin, Phys. Rev. B  59
  (1999) 7083.

\end{thebibliography}

\end{document}